\documentclass[usenatbib]{mn2e}
\usepackage{natbib}
\usepackage{epsfig}

\catcode`\@=11
\def\gta{\ifmmode{\,\mathrel{\mathpalette\@versim>\,}}
    \else{$\,\mathrel{\mathpalette\@versim>}\,$}\fi}
\def\lta{\ifmmode{\,\mathrel{\mathpalette\@versim<\,}}
    \else{$\,\mathrel{\mathpalette\@versim<}\,$}\fi}
\def\@versim#1#2{\lower 2.9truept \vbox{\baselineskip 0pt \lineskip
    0.5truept \ialign{$\m@th#1\hfil##\hfil$\crcr#2\crcr\sim\crcr}}}
\catcode`\@=12  
\def\figref#1{Fig.~\ref{#1}}
\def\kpc{\,{\rm kpc}}
\def\hi{H\,{\sc i}}
\def\WHIM{{\sc whim}}
\def\HVC{{\sc hvc}}

\def\msun{\,{\rm M}_\odot}
\def\avrhodyn{{\bar{\rho}_{\rm dyn}}}

\def\Tzero{{T_0}}
\def\nezero{n_{\rm e,0}}
\def\rzero{r_0}
\def\rhalf{r_{\rm half}}
\def\rprime{r^{\prime}}
\def\Phizero{{\Phi_0}}
\def\grad{{\nabla}}

\def\d{{\rm d}}
\def\i{{\rm i}}

\def\vv{{\bf v}}

\def\LX{L_{\rm X}}
\def\Zsun{Z_{\rm \odot}}
\def\Mgas{M_{\rm gas}}

\def\xvt{{\bf x}_{\rm t}}

\def\kb{k_{\rm B}}

\def\kt{k_{\rm t}}
\def\ktkcrit{(\kt/k)_{\rm crit}}
\def\mp{m_{\rm p}}
\def\Msun{{M}_{\odot}}
\let\msun=\Msun

\def\Tvir{T_{\rm vir}}
\def\tth{t_{\rm th}}
\def\tBV{t_{\rm BV}}
\def\tcool{t_{\rm cool}}
\def\omegaBV{\omega_{\rm BV}}
\def\omegac{\omega_{\rm c}}

\def\omegatilde{\tilde{\omega}}

\def\kr{k_{\rm r}}
\def\omegath{\omega_{\rm th}}
\def\tdyn{t_{\rm dyn}}

\def\Mdyn{{M}_{\rm dyn}}

\def\rhodyn{{\rho}_{\rm dyn}}
\def\lambdacrit{\lambda_{\rm crit}}

\def\ne{n_{\rm e}}
\def\ne2{n_{\rm e}^2}

\def\ne{n_{\rm e}}
\def\nez{n_{{\rm e},0}}

\def\cmmcube{\,{\rm cm}^{-3}}
\def\cm{\,{\rm cm}}
\def\kms{\,{\rm km}\,{\rm s}^{-1}}

\def\Gyr{\,{\rm Gyr}}
\def\sm1{\,{\rm s}^{-1}}
\def\kelvin{\,{\rm K}} \let\K=\kelvin
\def\log{\,{\rm log}}
\def\erg{\,{\rm erg}}
\def\ergs{\,{\rm erg}\,{\rm s}^{-1}}
\def\ergscm3{\,{\rm erg}\,{\rm s}^{-1}\,{\rm cm}^{-3}}
\def\kspitzer{\kappa_{\rm S}}
\def\gammap{{\gamma^{\prime}}}
\def\keV{\,{\rm keV}}
\def\nH{n_{\rm H}}
\newcommand{\ls}{{_<\atop^{\sim}}}

\def\xspec{{\sc xspec}}

\title[HVCs from thermal instability?]
{Do high-velocity clouds form by thermal instability?}

\author[J. Binney, C. Nipoti and F. Fraternali]{James Binney$^1$\thanks{E-mail: binney@thphys.ox.ac.uk}, 
Carlo Nipoti$^2$ and Filippo Fraternali$^2$ \\
$^{1}$ Rudolf Peierls Centre for Theoretical Physics, Keble Road, Oxford OX1 3NP, UK\\
$^{2}$Dipartimento di Astronomia, Universit\`a di Bologna, via Ranzani 1, 40127 Bologna, Italy\\}

\begin{document}

\date{Accepted 2009 May 20.  Received 2009 May 15; in original form 2009 February 25}

\pagerange{\pageref{firstpage}--\pageref{lastpage}} \pubyear{2009}

\maketitle

\label{firstpage}

\begin{abstract}
  We examine the proposal that the \hi\ ``high-velocity'' clouds (\HVC
  s) surrounding the Milky Way and other disc galaxies form by
  condensation of the hot galactic corona via thermal instability.
  Under the assumption that the galactic corona is well represented by
  a non-rotating, stratified atmosphere, we find that for this
  formation mechanism to work the corona must have an almost perfectly
  flat entropy profile.  In all other cases the growth of thermal
  perturbations is suppressed by a combination of buoyancy and thermal
  conduction.  Even if the entropy profile were nearly flat, cold
  clouds with sizes smaller than $10\kpc$ could form in the corona of
  the Milky Way only at radii larger than $100\kpc$, in contradiction
  with the determined distances of the largest \HVC\ complexes.
  Clouds with sizes of a few kpc can form in the inner halo only in
  low-mass systems. We conclude that unless even slow rotation
  qualitatively changes the dynamics of a corona, thermal instability
  is unlikely to be a viable mechanism for formation of cold clouds
  around disc galaxies.
\end{abstract}


\section{Introduction}

It is widely accepted that star-forming disc galaxies such as the
Milky Way have sustained their star formation over gigayears by
accreting gas at a fairly steady or slowly declining rate
\citep[e.g.][]{Chiappini01}. The source of this gas is uncertain.  The
Milky Way and other star-forming disc galaxies (such as M31) that have
been studied with sufficient sensitivity in the 21-cm
hyperfine-structure line of hydrogen are surrounded by clouds of
\hi\ \citep[e.g.][]{Wakker97,Westmeier05}, but the amount of gas
contained in these ``high-velocity'' clouds (\HVC s) is sufficient to
sustain accretion onto the disc for about a gigayear
\citep[e.g.][]{Wakker07,Sancisi08}. The only realistic source of gas
for sustained accretion onto discs is the warm-hot intergalactic
medium (\WHIM), which is believed to contain over half the baryons in
the Universe. This belief is based on a combination of big-bang
nucleosynthesis theory and observations of the cosmic microwave
background \citep[e.g.][]{Fukugita}, so within the framework of
standard cosmology it must be considered robust.  Empirical evidence
for the existence of the \WHIM\ includes the requirement for a medium
to confine clouds of interstellar gas far from the Galactic plane
\citep{Spitzer56}, the observation of CIV, OV and OVI absorption along
lines of sight that pass close to the Galaxy's \HVC s
\citep{FUSE,Fox06}, and the morphology of the Magellanic Stream and
individual \HVC s \citep{Putman,Bruns00}.

The relationship of the \WHIM\ to circum-galactic clouds of \hi\ is
unclear.  Undoubtedly much of the \hi\ that is observed more than a
kiloparsec from galactic planes has been pushed off the star-forming
plane by the mechanical feedback from star formation
\citep[e.g.][]{FraternaliB08} -- the galactic fountain effect
\citep{Shapiro76}.  \citet{Bregman80} proposed that the galactic
fountain model could explain also the formation of the \HVC s by
condensation from hot gas outflows from the disk, but
\citet{Ferrara92} argued on theoretical grounds that this mechanism is
unlikely to work. The hypothesis that HVCs formed by condensation of
the hot gas of the galactic fountain is discarded also on
observational grounds, because such gas cannot be less metal rich than
the local interstellar medium, whereas the high-velocity cloud Complex
C, which lies $6-11\kpc$ from the plane, has $Z\simeq0.15Z_\odot$
\citep{Wakker07}.  This finding suggests that some \HVC s, especially
those more than $\sim5\kpc$ from the plane, represent matter that is
entering the Galaxy for the first time \citep{Oort70}.

At least some \HVC s must have been stripped from infalling satellites
because some clouds are clearly associated with the Magellanic Stream,
which is itself thought to have been torn from the Small Magellanic
Cloud \citep{Putman}. Counts of stars in the Sloan Digital Sky Survey
\citep{Bell08} and in deep surveys of M31 \citep{MegaCam} have shown
that the outer stellar halos of the Galaxy and M31 are far from
relaxed and are best thought of as a superpositions of dissolving
satellites.  Any gas that belonged to these satellites is likely to
give rise to \HVC s.

However, the supply of cold gas from infalling satellites falls far
short of what a typical star-forming disc galaxy needs to sustain its
star formation \citep[e.g.][]{Sancisi08}.  In the long run that gas
{\it must\/} come from the \WHIM\ because the latter is the only gas
reservoir of sufficient capacity.  So if star-forming galaxies obtain
the gas they need by accreting \HVC s, these clouds must form from the
\WHIM.  A case for the formation of \HVC s from the \WHIM\ has been
argued on the basis of simulations of galaxy formation that follow the
baryons with SPH \citep{Kaufmann06,SommerLarsen06,Peek08,Kaufmann09}.
This case is not strong because (a) the mass scale of \HVC s is
smaller than the mass resolution of state of the art cosmological
simulations, (b) there are grave doubts about the reliability of SPH
when the fluid contains extremely steep density gradients such as
those expected at the interface of the \WHIM\ and an \hi\ cloud and
(c) multi-phase media in SPH simulations might be subject to
artificial overcooling \citep[][but see also Price
  2008]{Marri03,SPHtest,Agertz07}.  In this paper we use largely
analytic arguments to examine the likelihood that \hi\ clouds can
condense from the \WHIM\ $\gta10\kpc$ above the main \hi\ disc.

The dynamics of the \WHIM\ around a galaxy such as the Milky Way bears
a close resemblance to the dynamics of ``cooling flows'' in groups and
clusters of galaxies. The gas in a cooling flow is hotter and denser
than the \WHIM, so it can be readily studied through its X-ray
emission. When observations showed that the cooling time at the centre
of a typical cooling flow is significantly less than the Hubble time,
it was assumed that the gas was flowing towards the centre and there
generating a pool of cold gas from which stars formed
\citep{Silk,CowieB}. Shortly afterwards it became clear that this
picture was not viable because (a) it predicted a radial X-ray
surface-brightness profile that was too centrally concentrated, and
(b) insufficient young stars were present at small radii, no matter
what the initial mass function.  The hypothesis then took firm root
that $\sim100\kpc$ from the centre, thermal instability in the
inflowing gas caused clouds of cold gas to ``drop out''
\citep{Nulsen}.  These clouds were presumed to be too small to be
detected, but removed sufficient X-ray emitting gas from the cooling
flow to reconcile the theoretical and observational surface-brightness
profiles.  Eventually this theory of ``distributed mass drop out'' was
ruled out by the absence of spectral lines in soft X-rays that
should have accompanied the formation of the clouds \citep{Peterson}.
However, more than a decade before the decisive observational evidence
arrived the intellectual foundations of the theory had been shot away
by \citet[][hereafter MRB]{Malagoli}, who pointed out that a hot
atmosphere that is confined by a gravitational field and radially
stratified by specific entropy is not thermally unstable in the sense
of \cite{Field65}. Subsequent papers clarified important aspects of
the problem \citep[Balbus \& Soker 1989, hereafter
  BS,][]{Tribble89,Balbus91} but confirmed the basic physical
principle. Unfortunately, it seems that the community that is now
working on galaxy formation is unaware of this principle
\citep[e.g.][Peek et al.~2008]{MallerB}, and there is a danger that studies of
the \WHIM\ will fall into the same trap as did early studies of
cooling flows. In particular, condensation of the \WHIM\ into \HVC s
$\gta10\kpc$ above the \hi\ disc is an instance of distributed mass
drop out, albeit in a different parameter regime. In this paper we
examine this possibility critically.

In Section 2 we summarise the analytical work of MRB and BS, and
explain the underlying physics. In Section 3 we describe the simple
models of galactic coronae in which we study thermal
instability. Section 4 presents both the result of applying linear
theory to these models and studying the time evolution of spherical
model atmospheres.  Section 5 sums up and discusses the implications
of our results for galaxy accretion.

\section{Thermal instability in stratified coronae}

The \WHIM\ around a galaxy like the Milky Way is expected to be
approximately in equilibrium, stratified in the galactic gravitational
potential. The thermal instability in gravitationally stratified
coronae has been studied by several authors
\citep[e.g. MRB,][BS]{WhiteS,Tribble89}, mainly in the context of the
study of cooling flows in galaxy clusters. In many respects a galactic
corona is just a scaled-down version of the hot atmosphere of a galaxy
cluster. Thus, under the assumption that the corona is non-rotating
and spherically symmetric, the formalism of previous work of spherical
models of cooling flows can be applied straightforwardly. However, the
detailed thermal behaviour of the system is sensitive to the gas
temperature and density distribution, so the conclusions drawn in the
case of clusters cannot be a priori extended to lower-mass systems.

Here we follow the treatment of MRB who used Eulerian plane-wave
perturbations to study thermal instability in a spherically symmetric
stratified atmosphere in the presence of a flow, including the effects
of cooling and thermal conduction. It has been pointed out that when a
background flow is present, the use of Lagrangian perturbations is
preferable to that of Eulerian perturbations
\citep[][BS]{Balbus88,Tribble89}.  However, the main results of MRB
have been confirmed by the Lagrangian study of BS. Thus, for the
purpose of the present investigation we prefer to use the Eulerian
approach, which has the advantage of a simpler formalism than the
Lagrangian approach.

\subsection{Perturbation analysis}
\label{sec:per}

We summarise here the perturbation analysis of a non-rotating
spherical stratified atmosphere by MRB.  The system is governed by the
equations for mass, momentum and energy conservation
\begin{eqnarray}
{\partial \rho \over \partial t}+\grad\cdot(\rho\vv)&=&0,\label{eqmass}\\
{\partial \vv \over \partial t}+\vv\cdot\grad\vv&=&-{\grad p \over \rho}-\grad\Phi,\label{eqmom}\\
{p\over \gamma-1}\left[{\partial \over \partial
  t}+\vv\cdot\grad\right] \ln (p \rho^{-\gamma})&=&
\grad\cdot\left(f\kspitzer T^{5/2}{\bf \grad} T\right)\nonumber \\
&&\quad-\left({\rho\over \mu \mp}\right)^2\Lambda(T).\label{eqen}
\end{eqnarray}
Here $\Phi$ is the galactic gravitational potential, $\gamma=5/3$ is
the ratio of principal specific heats, $\Lambda$ is the cooling
function, $\mu$ is the mean gas particle mass in units of the proton
mass $\mp$, $\kspitzer\simeq1.84\times10^{-5}\erg \sm1
\cm^{-1}\kelvin^{-7/2}$ is Spitzer's coefficient of thermal
conductivity \citep[assuming a value 30 for the Coulomb
  logarithm;][]{Spi62} and $f\leq 1$ is the factor by which thermal
conduction is suppressed by a tangled magnetic field
\citep[e.g.][]{Bin81}. We are assuming that the thermal conductivity
is unsaturated, which will certainly be the case in an isothermal
corona since there the only temperature gradients are those associated
with perturbations and therefore of order $\delta T/\lambda$, where
$\delta T$ is infinitesimal and $\lambda$ is the perturbation's
wavelength. Near the edge of an adiabatic corona the temperature
gradient is steep and the density low (\figref{figadiab} below), so
the conductivity may saturate. Reduced conductivity will underline the
tendency of these regions to be thermally unstable.

  Apart from the presence of the factor $f$,  we
  treat the hot galactic gas as unmagnetized. We do not attempt to
  model the magnetic field in the corona for simplicity and because
  its properties are poorly constrained. An ordered magnetic
  field can have subtle
  effects on thermal instability \citep{Loewenstein1990,Balbus91}, but, as we
  explain in Section 5,  far from the
  disc the field is unlikely to be ordered.

The unperturbed corona is assumed to be close to hydrostatic and
thermal equilibrium in the sense that the system is approximately in a
steady state over the time scales of interest even in the presence of
radiative cooling and thermal conduction.  Thus, the corona can be
described by the time-independent spherically-symmetric pressure $p$,
density $\rho$, temperature $T$ and velocity field $\vv=v \hat{\bf
  e}_{r}$, which satisfy equations (\ref{eqmass}) to (\ref{eqen}) with
vanishing partial derivatives with respect to $t$.

We linearise equations~(\ref{eqmass}-\ref{eqen}) with Eulerian
perturbations of the form $F+\delta F\exp(-\i\omega t+\i\kr r + \i{\bf
  \kt} \cdot \xvt)$, where $\xvt$ is the tangential vector, $\kr$ is
the radial wave-number, $\kt\equiv|{\bf \kt}|$ is the tangential
wave-number and $|\delta F|\ll|F|$.  Under the assumption of short
wave-length and low-frequency perturbations, the linearised equations
reduce to the following dispersion relation:
\begin{equation}
\label{eqdisp} 
\omegatilde^2+\i\omegatilde(\omegac+\omegath)-{\kt^2\over k^2}\omegaBV^2=0,
\end{equation}
 where $\omegatilde\equiv\omega-\kr v$ is the frequency experienced by
 a perturbation that rides with the background flow and
 $k^2=\kr^2+\kt^2$.  In the equation above
\begin{eqnarray}
\omegac&\equiv&{\gamma-1\over\gamma}{k^2 f\kspitzer T^{7/2}\over p},\label{eqomegac} \\
\omegath&\equiv&-{\gamma-1\over\gamma}{n\Lambda(T)\Delta(T)\over
\kb T},
\label{eqomegath}
\end{eqnarray}
where $n=\rho/\mu\mp$ is the particle number density, 
\begin{equation}
\Delta(T)\equiv 2-{\d\ln \Lambda \over \d \ln
T},
\end{equation}
and $\omegaBV$ is Brunt-V\"ais\"al\"a (hereafter BV) frequency defined as
\begin{eqnarray}
\omegaBV&\equiv&\left\{{g\over T}\left[{\d T\over \d r} -\left({\d T\over \d r}\right)_{\rm adiab}\right]\right\}^{1/2}\nonumber\\
&=&\left[{g\over T}\left({\d T \over \d r} +{\gamma-1\over \gamma}{ g \mu \mp\over \kb}\right)\right]^{1/2},\label{eqomegabv} 
\end{eqnarray}
where $g=\Vert \grad\Phi\Vert$ and $(\d T/\d r)_{\rm adiab}$ is the
adiabatic temperature gradient.  The BV frequency $\omegaBV$ is
clearly real in the relevant case of convectively stable
configurations.

\begin{figure}
\includegraphics[width=\hsize]{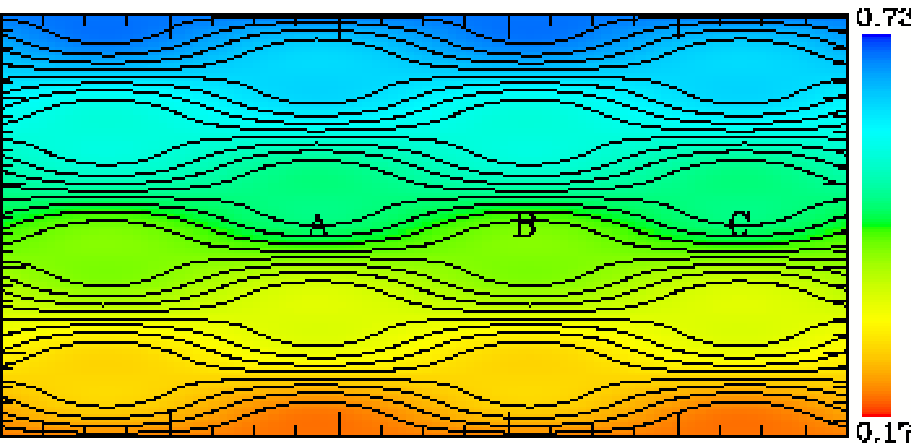}
\includegraphics[width=\hsize]{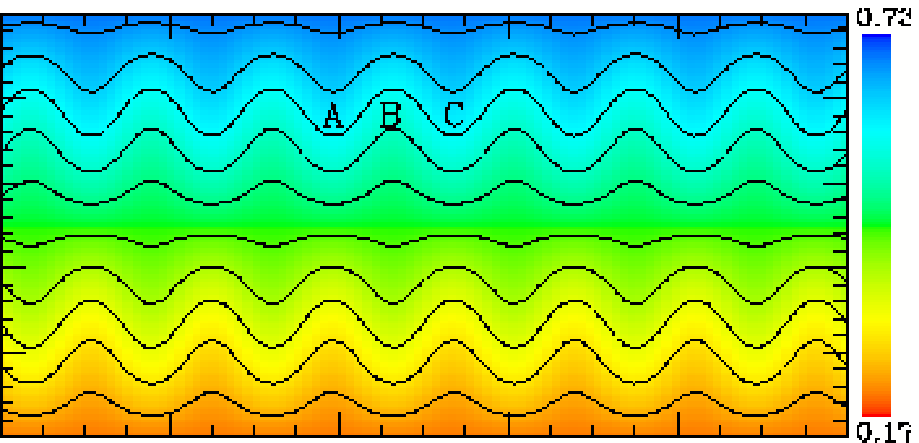}
\caption{Contours of constant specific entropy (in the plane of the
  tangential and radial coordinates) for a model corona with
  perturbations that have $\kt/\kr=2/7$ (upper panel) and $7/2$ (lower
  panel). The gravitational field is directed from top to
  bottom. Specific entropy is in arbitrary units.}
\label{fig:cartoon}
\end{figure}

In the absence of cooling and thermal conduction ($\omegac=\omegath=0$),
the dispersion relation (\ref{eqdisp}) yields 
 \begin{equation}\label{dispsimp}
\omegatilde={\kt\over k}\omegaBV.
\end{equation}
Thus $\omegaBV$ is the frequency at which radially-oriented needles of
gas bob up and down around their equilibrium radii. As $\kt/k$ falls
from unity, the shape of the oscillating region flattens through a
sphere, reaching a tangentially oriented disc in the limit $\kt/k\to0$
(\figref{fig:cartoon}).  According to equation~(\ref{dispsimp}), the
oscillation frequency declines from $\omegaBV$ to zero as the body is
flattened from a needle to a disc. To understand this result
physically, consider the upper panel of \figref{fig:cartoon}, showing
isoentropy contours for a model corona with perturbations that have
$\kt/\kr=2/7$. In the absence of cooling or conduction, specific
entropy is unchanged as the fluid moves, so the colours and contours
are frozen into the fluid. At the points marked A and C in the figure,
the entropy perturbation is positive because the fluid is displaced
downward from equilibrium, while at point B the fluid is displaced
upwards.  The upward movement of the contour above B has clearly been
enabled by fluid moving away horizontally towards the points A and
C. In the lower panel -- showing isoentropy contours for the same
model corona, but with $\kt/\kr=7/2$ -- it is clear that the upward
movement at B requires a less significant horizontal movement of
fluid. Consequently, the effective inertia that has to be overcome by
the perturbation's buoyancy increases with the flattening of the
perturbation, so $\omegatilde$ decreases with $\kt/k$.

\begin{figure}
\includegraphics[width=\hsize]{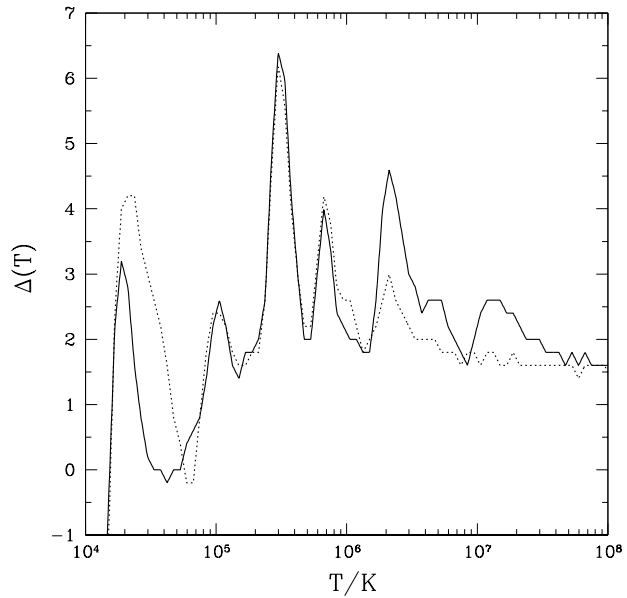}
\caption{The quantity $\Delta(T)=2-{\d\ln \Lambda / \d \ln T}$ as a
  function of temperature (full line: $Z=\Zsun$, dotted curve
  $Z=0.03\Zsun$) for the cooling function of
  \protect{\citet{SutherlandD}}.}
\label{fig:omegath}
\end{figure}

In the absence of conduction or a confining gravitational force
($\omegac=\omegaBV=0$) equation (\ref{eqdisp}) states that
$\omegatilde=-\i\omegath$, so the flow is unstable when $\omegath<0$
\citep{Field65}. From equation~(\ref{eqomegath}) $\omegath<0$ when
$\Delta(T)>0$, which is always the case at temperatures $\gta 7\times
10^4\K$, as is apparent from \figref{fig:omegath}, in which
$\Delta(T)$ is plotted for the tabulated cooling function of
\citet{SutherlandD}, which we adopt throughout the paper.  Thus, the
condition for thermal instability is satisfied in the temperature
range of interest and we can specialise to the case $\omegath<0$. The
conduction frequency $\omegac>0$ always, and conduction replaces the
Field instability with exponential decay when
$\omegac>|\omegath|$. Since $\omegac\propto k^2$, the
shortest-wavelength modes are always damped.

In the course of buoyant oscillations, a parcel of gas is half the
time overdense and cooler with respect to its surroundings, and half
the time underdense and warmer, so unless
$|\omegac+\omegath|\gta(\kt/k)\omegaBV$, neither conduction nor
radiation will have a big impact on buoyant oscillations. This
expectation is borne out by exact analysis of the dispersion relation
(\ref{eqdisp}), which shows that when $\omegac=0$ modes are
monotonically growing if $\kt/k$ is smaller than
\begin{equation}\label{eqktkcrit} 
\left({\kt\over k}\right)_{\rm
crit}\equiv{|\omegath|\over2\omegaBV}.
\end{equation}
In addition, thermal conduction stabilises perturbations with
wavelengths smaller than
 \begin{equation}
\label{eqlambdacrit} 
\lambdacrit\equiv {2\pi\over n}\left[{f\kspitzer T^{7/2}\over\Lambda(T)\Delta(T)}\right]^{1/2}.
\end{equation}

Although sufficiently flattened perturbations can be thermally
unstable, equation (\ref{eqlambdacrit}) places a lower bound on the
thickness of any thermally unstable disc, so such a disc's transverse
size must be considerable.  It is likely that the hot atmospheres of
galaxies, galaxy groups and clusters are turbulent by virtue of cosmic
infall, galactic winds and jets from AGN.  Consequently, an extended
disc-like overdensity is unlikely to remain flat and exactly
tangentially oriented for more than a dynamical time. Once turbulence
has distorted and/or rotated a perturbation, its value of
$\omegatilde$ will rise towards $\omegaBV$ and it will cease to be
thermally unstable. Therefore we can discount the possibility that
$\omegatilde\ll\omegaBV$.

\subsection{Physical implications}

\begin{figure}
\includegraphics[width=\hsize]{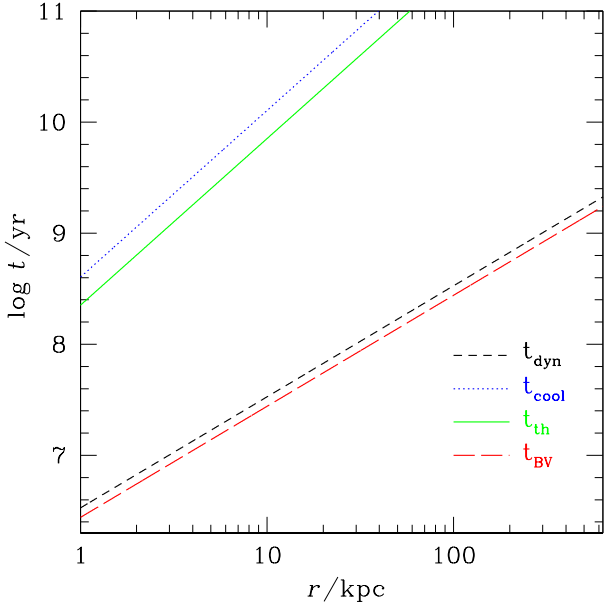}
\caption{The radial dependence of dynamical, cooling, thermal
  instability and BV timescales in a model atmosphere representative
  of those of massive elliptical galaxies \citep[model HM of][see also
    Table~\ref{GalTab}]{NipotiB07}.}\label{fig:hm}
\end{figure}

The thermal and dynamical evolution of a galactic corona is ultimately
determined by four timescales: the dynamical time of the system
$\tdyn$, the gas isobaric cooling time
\begin{equation}
  \label{eqtcool} 
  \tcool\equiv{\gamma \over \gamma-1} 
{\kb T \over  n \Lambda(T)},
\end{equation}
the BV time $\tBV\equiv\omegaBV^{-1}$ and the thermal instability time
$\tth\equiv|\omegath^{-1}|$. Figure~\ref{fig:hm} illustrates this by
plotting these timescales for a particular atmosphere representative
of those of massive elliptical galaxies \citep[model HM of][with
  metallicity $Z=0.3\Zsun$; see also Table~\ref{GalTab}]{NipotiB07}.
Clearly, a necessary condition for a corona to be in equilibrium is
$\tdyn<\tcool$, so in \figref{fig:hm} the dotted line lies above the
short-dashed line: a plasma in which this condition is violated is
experiencing a cooling catastrophe, and the question of its thermal
stability is meaningless. Let us focus on systems with $\tdyn<\tcool$,
which are approximately in equilibrium and, at least in principle, can
be prone to thermal instability. As we have seen, thermal
perturbations of average shape ($\kt \sim \kr$) can grow only if
$\tth<\tBV$, because otherwise buoyancy is effective in restoring the
unperturbed entropy profile. In the typical case of subadiabatic
temperature profile, the BV time is of the order of the dynamical time
(both being determined mainly by the gravitational field), while the
thermal instability time is of the order of the cooling time (both
being determined mainly by the cooling function).  Hence in
\figref{fig:hm} the long-dashed line lies close to the short-dashed
line, while the solid line lies close to the dotted line, which
necessarily lies above the short-dashed line. As a consequence, it is
not easy to find systems in which both the conditions $\tdyn<\tcool$
and $\tth<\tBV$ are satisfied, and this is the main reason why
stratified atmospheres are typically thermally stable as MRB pointed
out in the context of rich galaxy clusters.

However, from equations (\ref{eqomegath}) and (\ref{eqtcool}), 
\begin{equation}
\label{eqratio} 
{\tcool\over\tth}=|\Delta(T)|, 
\end{equation}
and \figref{fig:omegath} shows that $\tth<\tcool$ for $T\gta 8\times
10^4\K$. Consequently, for certain temperature and density
distributions within the system, $\tBV$ can be longer than $\tth$, and
$\tdyn$ shorter than $\tcool$, even when
$\tBV\sim\tdyn$. Figure~\ref{fig:omegath} shows that the scope for
simultaneously satisfying both conditions is largest at temperatures
$\sim3\times10^5\K$ where the cooling function increases steeply for
decreasing temperature.  Thus, the development of thermal instability
in galactic coronae, which have virial temperatures in this range,
cannot be excluded a priori.

As the temperature profile approaches the adiabatic temperature
profile, $\tBV$ becomes arbitrarily long (see
equation~\ref{eqomegabv}) -- in an adiabatic atmosphere there is no
tendency for buoyancy to restore displaced material to its original
radius. Hence an atmosphere with a very shallow entropy profile may
well be thermally unstable. Information on the entropy profile of
galactic coronae is then crucial for the issue of their thermal
stability (see also BS).

Finally, thermal conductivity can potentially give an important
contribution to the suppression of small-scale perturbations:
conductivity damps all perturbations with wavelength smaller than the
critical wavelength $\lambdacrit$ (equation~\ref{eqlambdacrit}). The
extent of this damping depends on the poorly constrained suppression
of conductivity, which we parametrise with the dimensionless factor
$f$. However, we will see that even strongly suppressed thermal
conduction can have non-negligible stabilising effect.

The discussion above leads to the conclusion that the window for the
development of thermal instability in galactic coronae is narrow, but
not necessarily closed. Thus, to address the question of the thermal
stability of the \WHIM\ it is necessary to apply the results of the
perturbation analysis to specific models of galactic coronae.

\section{Applications to galactic coronae}
\label{sec:app}

Application of the thermal instability analysis to the X-ray emitting
hot atmospheres of massive elliptical galaxies and clusters of
galaxies has shown clearly that these systems cannot develop thermal
instabilities (MRB, BS). This result is apparent from
Fig.~\ref{fig:hm}, which shows the relevant timescales for a massive
elliptical galaxy model: in this case the BV timescale is at least two
orders of magnitude shorter than the thermal instability timescale.

Here we address directly the question of whether the
\WHIM\ surrounding disc galaxies is prone to thermal
instability. Unfortunately, the properties of the coronae of disc
galaxies are not as well known as those of their more massive
counterparts, mainly because they have low X-ray surface brightnesses
on account of their low gas densities.  Despite various attempts to
detect in X-rays the coronae of disc galaxies, so far only upper
limits on their X-ray luminosity have been obtained
\citep[e.g.][]{Rasmussen09}.  In the cases in which extra-planar X-ray
emission is detected, it appears clumpy, with clear connections with
the star-forming regions, and is likely to be explained by galactic
fountains \citep[e.g.][]{Strickland, Li}. But also from these data we
can derive upper limits to the total luminosity of the hot halos.  In
the special case of the Milky Way different pieces of information can
be used to constrain the physical properties of the corona
\citep[e.g.][]{FukugitaP}.

Here we consider simple non-rotating spherical models of coronae
representative of the \WHIM\ surrounding specific disc galaxies, whose
gross properties are consistent with the observational
constraints. The neglect of rotation is a serious limitation of this
work because rotation must become dynamically important sufficiently
close to the disc if the corona is to feed the disc.  Results for
non-rotating coronae are of interest because the clouds we wish to
understand lie far from the disc where rotation is expected to be a
small effect dynamically, and linear analysis of a differentially
rotating coronae is a challenging problem.

The hot gas is in hydrostatic equilibrium with polytropic distribution
 $p(r)\propto[\rho(r)]^{\gammap}$, where $\gammap$ is the polytropic index:
the gas temperature profile is
\begin{equation}\label{TgammaPhi}
{T(r)\over \Tzero}=1-{\gammap-1 \over \gammap}{\mu\mp\over \kb\Tzero}
(\Phi-\Phizero),
\end{equation}
where $\Tzero\equiv T(\rzero)$, $\Phizero\equiv \Phi(\rzero)$ and
$\rzero$ is a reference radius -- in the following we assume
$\rzero=10\kpc$.  When $\gammap>1$ the electron number density
profile is
\begin{equation}
{\ne \over \nezero}=\left(T\over \Tzero\right)^{1/(\gammap-1)},
\end{equation}
where $\nezero\equiv\ne(\rzero)$ and $\ne=0.52n$ (assuming abundances
by mass $Y=0.25$, $X=0.75$).  When $\gammap=1$ the distribution is
isothermal at temperature $\Tzero$ and the electron density profile is
\begin{equation}
\ne(r)=\nezero\exp\left[-{\mu\mp\over\kb\Tzero}(\Phi-\Phizero)\right].
\end{equation}
The gravitational potential $\Phi$ is determined by the total mass
density distribution $\rhodyn$, so $\nabla^2\Phi=4\pi G\rhodyn$. The
dynamical time is
\begin{equation}
\label{eqtdyn}
\tdyn(r) \equiv\sqrt{3 \pi \over 16 G \avrhodyn}={\pi \over 2} {r^{3/2} \over \sqrt{G \Mdyn(r)}},
\end{equation}
where $\Mdyn(r)=4\pi\int_0^r \rhodyn(\rprime){\rprime}^2 \d \rprime$
is the dynamical mass of the system and $\avrhodyn=3\Mdyn(r)/ 4 \pi
r^3$ is the average mass density within $r$.

We assume here that the total density distributions of our model
galaxies follow the singular isothermal sphere
 \begin{equation}
\label{eqrhodyn}
\rhodyn(r) = {\kb \Tvir \over 2 \pi G \mu  \mp r^2},
\end{equation}
and that the gravitational potential is
\begin{equation}
\label{eqPhi}
\Phi(r) = 2 {\kb \Tvir \over\mu
  \mp} \ln\left({r \over \rzero}\right)+\Phizero,
\end{equation}
where $\Tvir$ is the virial temperature, which can be linked to the
galaxy's circular speed $v_{\rm c}$:
\begin{equation}
\label{eqTvir}
\Tvir = \frac{\mu \mp}{2 \kb} v_{\rm c}^2.
\end{equation}
When the gas is isothermal ($\gammap=1$) in equilibrium in the
potential of a singular isothermal sphere, the gas density
distribution reduces to the simple power law
\begin{equation}
\ne(r)=\nezero\left({r\over\rzero}\right)^{-\alpha},
\end{equation}
where $\alpha=2\Tvir/\Tzero$. 

In summary, each galaxy model is characterised by five parameters:
$\Tvir$, $\gammap$, $\Tzero$, $\nezero$ and the metallicity $Z$, which
enters the calculation of the cooling function. 

\subsection{Observational constraints}

We use three galaxies as raw models for three categories of galaxies
with different masses.  The first is NGC\,5746, a very massive spiral
with flat rotational speed $\simeq310 \kms$ \citep[corresponding to a
  virial temperature $\Tvir\simeq 3.4\times 10^6$ K;][]{Rand08}, for
which we take from \citet{Rasmussen09} the upper limit on the X-ray
luminosity of the corona $\LX < 4 \times 10^{39} \ergs$.  The second
is the Milky Way, whose corona is constrained observationally as in
\citet{FukugitaP}. The third galaxy is NGC\,6503, a low-mass spiral
seen at an inclination of about 75 degrees, for which {\it Chandra\/}
observations yield an upper limit $\LX < 1.6 \times 10^{38} \ergs$ to
the X-ray halo luminosity \citep{Strickland}.  For this galaxy we use
$v_{\rm c} = 120 \kms$ \citep{Begeman87}, which by
equation~(\ref{eqTvir}) leads to a virial temperature $\Tvir = 5.1
\times 10^5 \K$.  In both NGC\,5746 and in NGC\,6503 a component of
extraplanar neutral gas has been detected \citep{Rand08, Greisen09}.

\begin{table*}
\centering
\caption{Parameters of the model galaxies. $\Tvir$: virial
  temperature. $\gammap$: polytropic index. $\Tzero$: electron
  temperature at $r=10\kpc$. $\nezero$: electron number density at
  $r=10\kpc$.  $Z$: metallicity. $\LX$: X-ray luminosity. $\Mgas$:
  corona gas mass. Isothermal and adiabatic disc galaxy models have
  names ending with ``i'' and ``a'', respectively. HM is an isothermal
  corona model representative of a high-mass elliptical galaxy.
  \label{GalTab}}
  \begin{tabular}{ccccccccc}               
Model      & $\Tvir$& $\gammap$ &$\Tzero$ & $\nezero$ & $Z$ & $\LX(1<r<10\kpc)$ & $\LX(10<r<100\kpc)$ & $\Mgas(r<350\kpc)$ \\
           & $(10^6\kelvin)$ & &$(10^6\kelvin)$& $(10^{-3}\cmmcube)$& $(\Zsun)$ & $(10^{39}\erg\sm1)$ & $(10^{39}\erg\sm1)$ & $(10^{11}\Msun)$ \\
\hline     
HM         &  $7.5$ &   1 & $10$  &  $3.5$ &  $0.3$   &  91.56 & 91.56 & 1.7 \\
N5746i   &  $3.4$ &   1 & $5.4$ &  $1.2$&  $0.03$  &   2.11 & 6.67 & 1.22 \\
MWi     &  $1.4$ &   1 & $1.9$ &  $2.6$ &  $0.03$  &   4.64 &  5.33 & 1.36 \\
N6503i   &  $0.51$&   1 & $0.82$&  $1$ &  $0.03$  &   0.04 & 0.13  & 1.01 \\
N5746a  &  $3.4$ & 5/3 & $9.7$ &  $0.62$&  $0.03$  &   0.27 &  12.88 & 1.23 \\
MWa    &  $1.4$ & 5/3 & $4$   &  $0.68$ &  $0.03$  &   0.19 &  5.58 &  1.30 \\
N6503a  &  $0.51$& 5/3 & $1.45$&  $0.5$ &  $0.03$  &   0.03 &  0.22 &  0.96 \\
\hline
\end{tabular}
\end{table*}

For each galaxy we consider an isothermal ($\gammap=1$) and an
adiabatic corona model ($\gammap=\gamma$). We verify that the models
are consistent with observational constraints by computing their X-ray
luminosities using the Mekal recipe \citep{Liedahl95} embedded in
\xspec.  The $0.3-2 \keV$ fluxes are calculated for a range of gas
temperatures ($0.1 <T< 1.5 \keV$) and for different metallicities
($Z$).  Given $f_{\rm \Delta E} (T,Z)$ the computed specific flux,
where $\Delta E = 0.3-2 \keV$, the plasma luminosity can be derived as
\begin{equation}
\label{mekal}
L_{\rm \Delta E} (T, Z) = 10^{-14} \int\d V\, \ne \nH f_{\rm \Delta E} (T, Z) ~\ergscm3
\end{equation}
where $\ne$ is the electron density and $\nH$ is the hydrogen density.
The factor in front the r.h.s.\ derives from the \xspec\
normalization.  For isothermal models the temperature in
equation~(\ref{mekal}) is constant with $r$ and so is $f_{\rm \Delta
  E}(T,Z)$, so one only integrates $\ne^2(r)$, assuming a constant
$\nH/\ne$ ratio. For adiabatic models, the temperature varies with $r$
and $f_{\rm \Delta E}(T,Z)$ has to be computed at each $r$.

The isothermal model of NGC\,5746 has total halo luminosity $\LX =3.99
\times 10^{39} \ergs$ in the range $5 < r < 40\kpc$, consistent with
the upper limit on the X-ray luminosity \citep{Rasmussen09}.  Our
isothermal model of the Milky Way is very similar to ``model 2'' of
\citet{FukugitaP}.  We checked that the obtained luminosity is in
agreement with observational constraints on galaxies similar to the
Milky Way, for instance NGC\,891 \citep{Strickland}, which happens to
be the case.  For the isothermal model of NGC\,6503, we assume
$\nez=1\times 10^{-3} \cm$, which gives a luminosity that matches the
upper limit by \citet{Strickland}. Similarly, the adiabatic models of
the three galaxies are constructed so that their X-ray luminosities
are consistent with observations.

\begin{figure*}
\centering
\includegraphics[width=\textwidth]{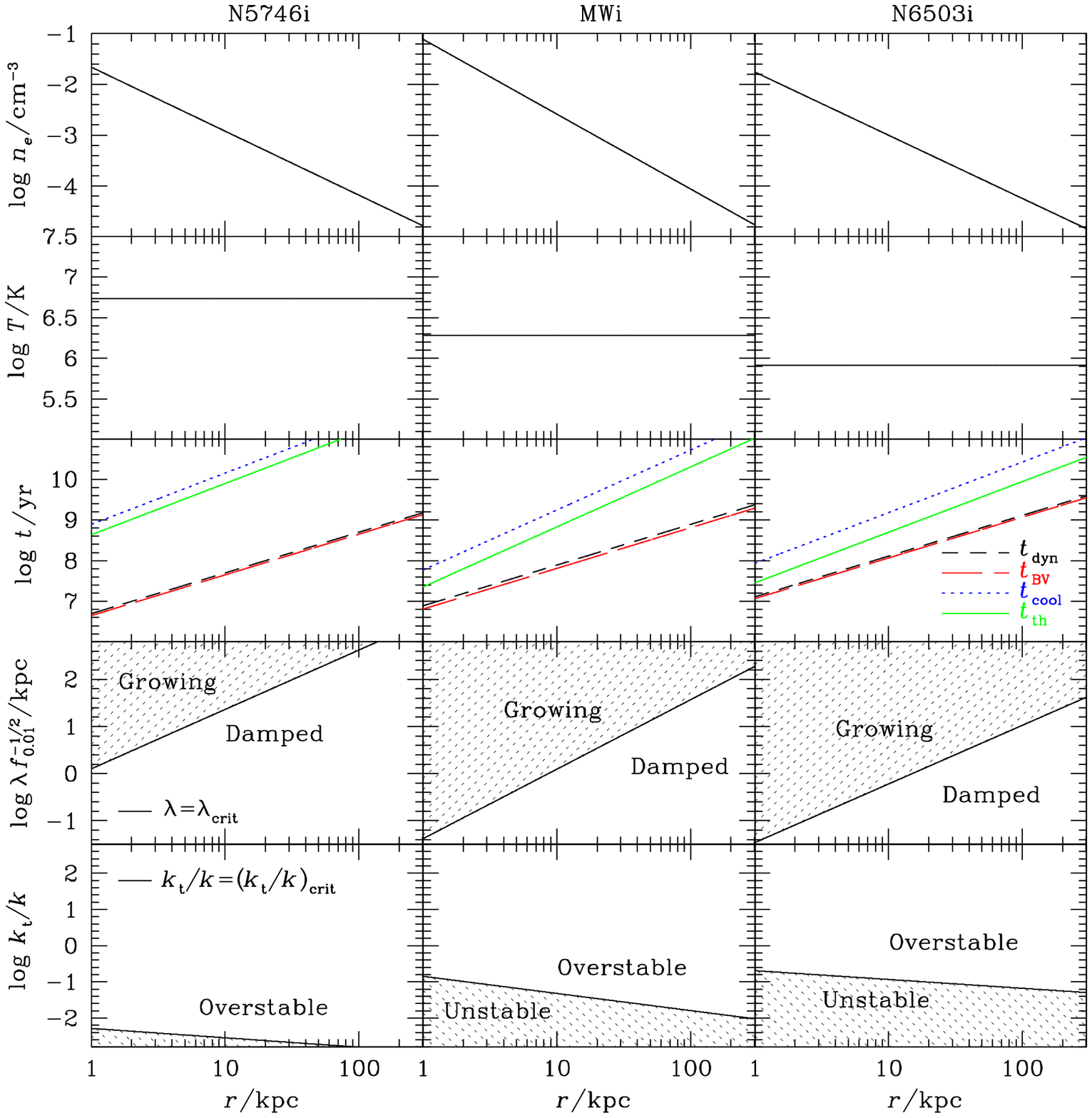}\\
\caption{From top to bottom: electron density, temperature,
  timescales, perturbation wavelength, and tangential-to-total
  perturbation wavenumber ratio as functions of radius for the
  isothermal models N5746i (left-hand column), MWi (central column)
  and N6503i (right-hand column). The timescales panels plot the
  dynamical (short dashed), BV (long dashed), cooling (dotted) and
  thermal instability (solid) timescales. In the wavelength panels the
  solid curves represent the critical wavelength $\lambdacrit$ for
  stabilisation by thermal conduction: only perturbations with
  wavelengths in the shaded areas grow ($f$ is the factor by which
  thermal conduction is suppressed; see equation~\ref{eqen}). In the
  wavenumber ratio panels the solid curves represent the critical
  wavenumber ratio $\ktkcrit$ for instability: only perturbations with
  ratio in the shaded areas are unstable.}
\label{figisot}
\end{figure*}

\begin{figure*}
\centering

\psfig{figure=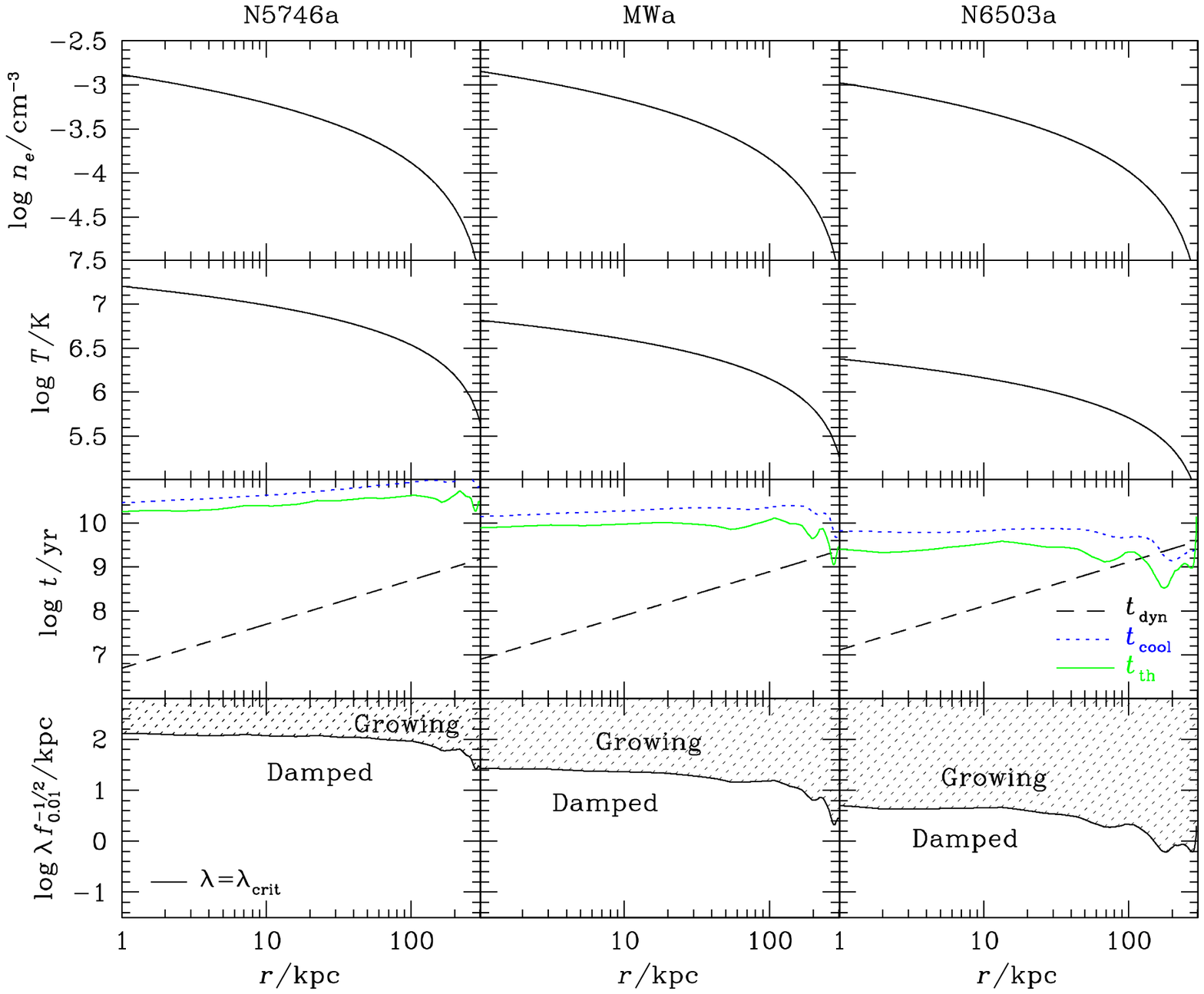,width=\hsize,angle=0,bbllx=18bp,bblly=260bp,bburx=577bp,bbury=725bp,clip=}
\caption{Same as Fig.~\ref{figisot} but for the adiabatic models
  N5746a (left-hand column), MWa (central column) and N6503a
  (right-hand column). In these cases $\tBV$ and $\ktkcrit$ are
  infinite at all radii, so we do not plot diagrams of $\kt/k$.}
\label{figadiab}
\end{figure*}

In contrast to the isothermal case, an adiabatic atmosphere has a
sharp edge and a well defined mass.  The location of the corona's edge
and thus the total gas mass depend strongly on the reference gas
temperature $\Tzero$.  We find that using the same $\Tzero$
(temperature at $r=10 \kpc$) as the isothermal models leads to gas
masses which are one or two orders of magnitude lower than what is
expected for a \WHIM\ corona.  Lacking constraints on the temperature
profiles, for the Milky Way corona we choose $\Tzero$ and $\nezero$ as
the largest values (in order to increase the mass of the corona) that
do not conflict with the luminosity constraints: these values, shown
in Table~\ref{GalTab}, give a total \WHIM\ mass of
$\sim1.3\times10^{11}\msun$, in line with the cosmological
predictions.  For NGC\,5746 and NGC\,6503 we keep the same ratio
between $\Tzero$ and $\Tvir$ as for the adiabatic Milky Way model
(i.e.\ same cut-off radius $\sim 350\kpc$), maximise $\nezero$
according to the X-ray luminosity constraints and find total masses of
the same order (see rightmost column of Table~\ref{GalTab}).  The
density and temperature profiles of our models are shown,
respectively, in the first and second rows of panels of
Fig.~\ref{figisot} (isothermal models) and Fig.~\ref{figadiab}
(adiabatic models).  The values of the parameters for all models are
summarised in Table~\ref{GalTab}.

\section{Results}

\subsection{Isothermal coronae}
\label{secisot}

Consider first coronae with isothermal gas distributions
($\gammap=1$), which are known to provide reasonable approximations to
the hot atmospheres of elliptical galaxies and galaxy clusters.  In
Fig.~\ref{figisot} we plot the results of applying the instability
analysis of Section~\ref{sec:per} to isothermal models of NGC\,5746
(model N5746i; left-hand column), the Milky Way (model MWi; central
column) and NGC\,6503 (model N6503i; right-hand column). The panels in
the third row of Fig.~\ref{figisot} are the analogues of
Fig.~\ref{fig:hm} and show as functions of radius the dynamical,
cooling, thermal instability and BV timescales. We see that in all
cases $\tBV$ (long-dashed line) is shorter than $\tth$ (solid line),
and the ratio $\tth/\tBV$ is higher in galaxies with higher virial
temperatures.  Hence in all cases the growth of thermal perturbations
is effectively countered by buoyancy.

A more detailed analysis of the behaviour of the perturbations can be
obtained from the bottom row of panels in Fig.~\ref{figisot}, which
show as a solid line the maximum wavenumber ratio $\ktkcrit$ a
perturbation can have and still grow monotonically: more tangentially
elongated modes [those with $\kt/k>\ktkcrit$] are overstable.  In all
cases $\ktkcrit$ is significantly smaller than unity: this means that
it is impossible for such an atmosphere to form blobs (i.e.,
perturbations with $\kr\sim\kt$ or $\kt/k\sim 1/\sqrt{2}$) by thermal
instability.  Consistent with the behaviour of the timescales,
$\ktkcrit$ increases for decreasing virial temperature of the system,
being $\lta 0.007$ in model N5746i and $0.06-0.25$ in model N6503i.
Thus, very tangentially elongated disturbances can in principle grow,
but we have seen that we do not expect them to grow in practice (see
Section~\ref{sec:per}).  The next-to-bottom row of Fig.~\ref{figisot}
quantifies ability of thermal conduction to stop growth by showing the
critical wavelength $\lambdacrit$ as a function of radius (solid
line).  Studies of galaxy clusters indicate that a realistic value of
the thermal conduction suppression factor is $f\sim0.01$ \citep[][and
  references therein]{NipotiB04}, and in this case at a radius $r$ all
perturbations smaller than $\sim 0.1 r$ in extent are damped in the
Milky-Way model MWi.  Stabilisation by thermal conduction is
proportionally more effective in higher-temperature systems than in
lower-temperature systems.  For instance, in NGC\,5746 at all radii
perturbations with wave-length smaller than the system's radius are
damped for $f=0.01$, while this is not the case in NGC\,6503.

We conclude that if galactic coronae are isothermal, the growth of
thermal perturbations is effectively suppressed by buoyancy and
thermal conduction. Suppression is particularly strong in very massive
systems such as NGC\,5746, but it is very effective also in coronae of
Milky-Way--like galaxies and smaller.  All these systems are unlikely
to form cold clouds by thermal instability.

\subsection{Polytropic coronae}

Though isothermality is a well motivated assumption, we cannot exclude
the possibility that galactic coronae have shallower specific entropy
profiles, so in this section we investigate coronae in which the
polytropic relation $p(r)\propto[\rho(r)]^{\gammap}$ holds between
pressure and density. As $\gammap$ ranges from unity to
$\gamma=\frac53$, these coronae range from isothermal to adiabatic.

The results for the adiabatic models (N5746a, MWa and N6503a in
Table~1) are shown in Fig.~\ref{figadiab}.  The cooling and thermal
instability timescales are now not power-law functions of radius
because density and temperature are no longer power-law functions as
in the isothermal models (see first and second rows of panels).  In an
adiabatic atmosphere there is no buoyancy to counter the growth of
thermal instability, so all modes with $\lambda>\lambdacrit$ grow
monotonically, independent of their wavenumber ratio [$\tBV\to\infty$
  and $\ktkcrit\to\infty$]. Thus the thermal stability of the gas is
determined by the critical wavelength $\lambdacrit$, which is shown in
the bottom row of panels in Fig.~\ref{figadiab}.  If the thermal
conduction suppression factor is $f=0.01$, perturbations are
effectively stabilised in model N5746a, but not as much in models MWa
and N6503a.  Thus, thermal instabilities cannot develop in the corona
of NGC~5746, even if it has an adiabatic distribution of gas. In the
case of the Milky Way, perturbations smaller than $\sim10\kpc$ in size
are expected to be stabilised by thermal conduction for $r
\ls200\kpc$.  In the case of NGC\,6503 perturbations smaller than
$2.5\kpc$ in size are stable for $r \ls 100\kpc$. However, if the
coronae of lower-mass systems such as the Milky Way and NGC\,6503 have
adiabatic distributions of gas, the growth of sufficiently large
perturbations by thermal instability is not excluded at large radii.
The rightmost panel in the third row of \figref{figadiab} indicates
that at $r\gta 150\kpc$ in model N6503a the perturbation analysis does
not apply because at these radii $\tcool\lta\tdyn$ so the corona is
not in hydrostatic equilibrium; catastrophic cooling is occurring.

\begin{figure}
\centering
\includegraphics[width=0.5\textwidth]{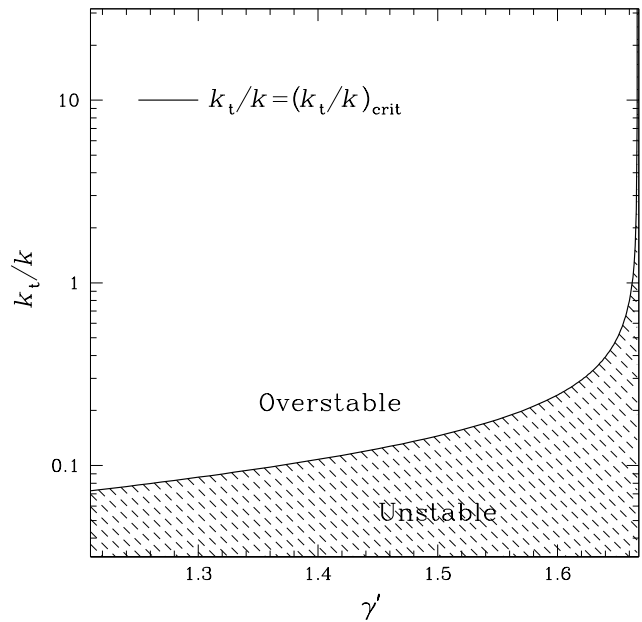}\\
\caption{Critical wavenumber ratio $(\kt/k)_{\rm crit}$ as a function
  of the polytropic index $\gammap$ at the half-mass radius $\rhalf$
  for Milky-Way--like polytropic models with $\Tvir=1.4\times10^6\K$
  and $Z=0.03\Zsun$.}
\label{figpoly}
\end{figure}

The assumption of perfectly adiabatic atmosphere is quite extreme, so
it is natural to explore the thermal instability of coronae with
entropy profiles intermediate between the steep ones of isothermal
models and the perfectly flat ones of adiabatic models. For this
purpose we consider polytropic models with $1<\gammap<\gamma$.  For a
polytropic distribution with index $\gammap$ the BV frequency is
 \begin{equation}
\omegaBV^2 ={g^2\over T}{\mu \mp\over \kb}[h(\gamma,\gammap)]^2,
\end{equation}
where
\begin{equation}
h(\gamma,\gammap)\equiv\left({1\over \gammap}-{1\over \gamma}\right)^{1/2}. 
\end{equation}
Thus we obtain the following expression for the maximum ratio of
wavenumbers that is consistent with  growing mode:
\begin{equation}
\label{eqktkcritpol} 
\left({\kt\over k}\right)_{\rm crit}=
{\gamma-1\over2\gamma h(\gamma,\gammap)}
{n \Lambda(T)\Delta(T)
\over
(\mu \mp \kb)^{1/2} T^{1/2} g}.
\end{equation}
We focus on Milky-Way--like polytropic models of index $\gammap$, with
$\Tvir=1.4\times10^6\K$, $Z=0.03\Zsun$, and values of $\nezero$ and
$\Tzero$ such that they have the same gas mass within $500\kpc$ and
the same value of the specific entropy at the radius $\rhalf$
containing half of the gas mass as the adiabatic model MWa. In
Fig.~\ref{figpoly} we plot as a solid line $\ktkcrit$ as a function of
$\gammap$ at $r=\rhalf$, which is in the range $140-170\kpc$ for these
models. Only perturbations with $\kt/k<\ktkcrit$ (shaded area in the
diagram) are unstable. We see that as soon as $\gammap$ deviates from
$\gamma$, $\ktkcrit$ drops from infinity to values smaller than unity,
and $\ktkcrit\lta 0.2$ for $\gammap\lta 1.55$. This means that thermal
instability is efficiently countered by buoyancy even in polytropic
distributions with quite shallow entropy profiles. We conclude that
unless the Milky Way corona is very close to adiabatic, thermal
perturbations cannot grow.

\subsection{Time evolution of polytropic coronae}\label{sec:timeevol}

\begin{figure*}
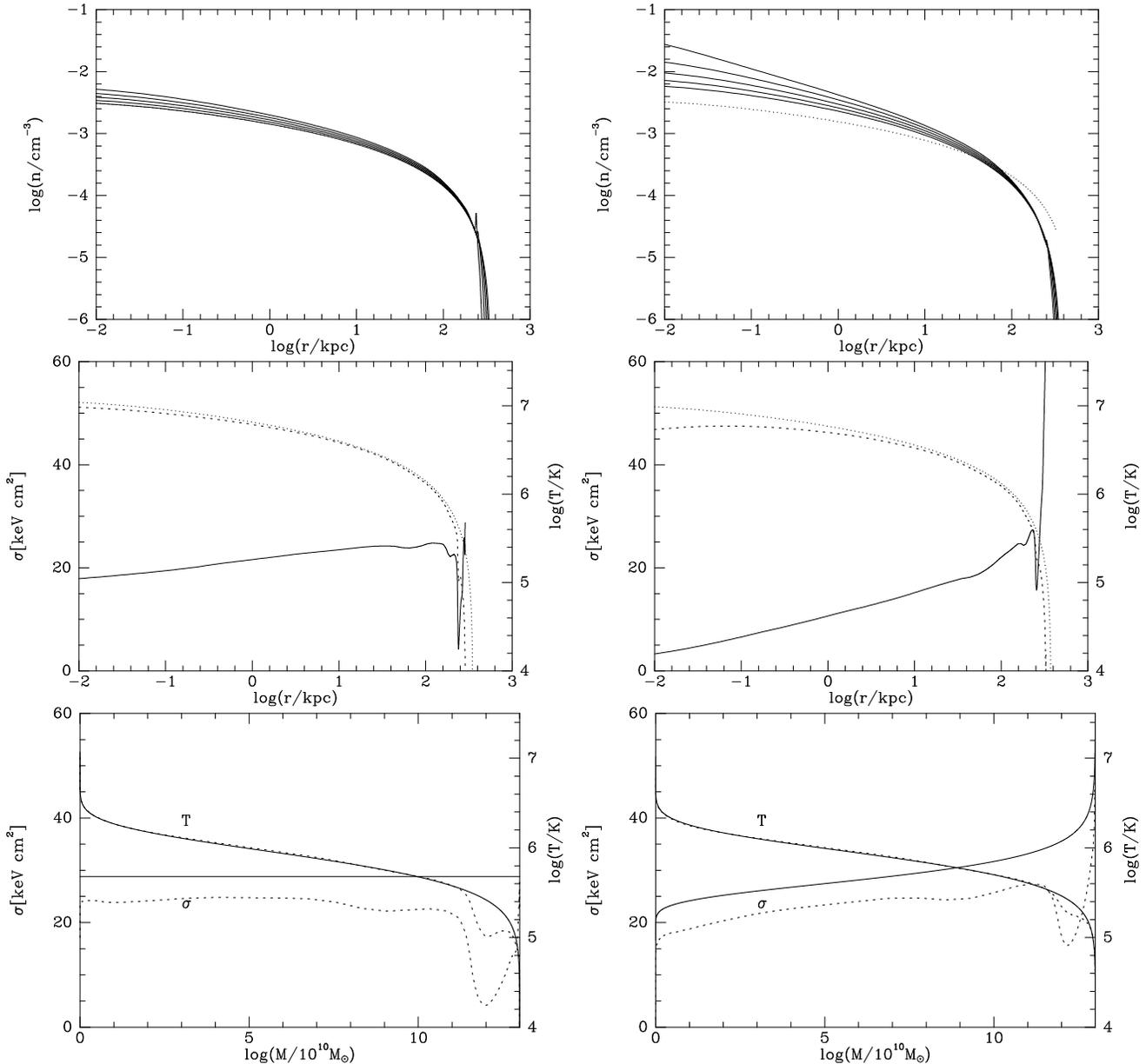

\centerline{\includegraphics[width=.44\hsize]{adiabdens.ps}\quad\qquad
  \includegraphics[width=.44\hsize]{polydens.ps}\quad}
\centerline{\includegraphics[width=.48\hsize]{adiabr.ps}\quad
  \includegraphics[width=.48\hsize]{polyr.ps}}
\centerline{\includegraphics[width=.48\hsize]{adiabT.ps}\quad
  \includegraphics[width=.48\hsize]{polyT.ps}}
 \caption{Evolution over $1\Gyr$ of model coronae for the Milky
   Way. Left column: an initially adiabatic model
   ($\gammap=1.6667$). Right column: a model with $\gammap=1.55$. The
   full curves in the top panels show the increase in the central
   electron density over $1\Gyr$, while the dotted curve in the top
   right panel shows the initial density profile of the adiabatic
   model of the left column. In the central and bottom panels the left
   scale gives $\sigma=p\rho^{-\gamma}$ (in units of $\keV \cm^{2}$ )
   and the right scale gives $\log\,T$. In the central panels the
   independent variable is radius and broken curves show temperature
   (the dotted curve shows the initial condition). In the bottom
   panels the independent variable is mass shell and initial and final
   conditions are shown by full and dashed curves respectively. The
   coronae were evolved by the technique described in Kaiser \&
   Binney~(2003) with metallicity $Z=0.03\Zsun$.}
 \label{KBfig}
\end{figure*}

In the absence of strong observational constraints, we try here to
address on theoretical grounds the question of whether the coronae of
disk galaxies are close adiabatic by exploring how a corona with
a shallow entropy profile evolves in time in the presence of cooling.  If
a steep outward-increasing entropy profile were produced quickly, we
could discard the hypothesis that coronae have very shallow entropy
profiles. For this purpose, we consider the extreme case of a corona
that starts from a perfectly flat entropy profile ($\gammap=\gamma$
polytrope) and the more general case of coronae with shallow outward
increasing entropy profiles ($\gammap<\gamma$ polytropes).

The left panels in \figref{KBfig} show how an initially adiabatic
model of the corona of the Milky Way evolves over $1\Gyr$. Although
the central density rises steadily (top left), the temperature profile
changes significantly only at the outer edge of the corona (bottom
left). Here the temperature was initially $\lta3\times10^5\K$, and, as
one moves outwards, the increase in the efficiency of cooling with
falling temperature outweighs the reduction in efficiency with
decreasing density. Consequently, the initially flat profile of
specific entropy $\sigma$ starts to fall with radius. In this region
of outwards-decreasing specific entropy, the corona is convectively
unstable and clouds of cool gas will fall inwards. Further in, the
specific entropy profile remains extremely flat as entropy is radiated
away, and regions of depressed temperature will not be stabilised by
buoyancy.  Thus a non-linear computation confirms that in the extreme
case of an adiabatic atmosphere perturbations large enough not to be
damped by thermal conduction can form cool clouds.

It is unlikely that all the corona's gas will have exactly the same
specific entropy -- initially some gas is bound to have more specific
entropy than other gas, and the galaxy's gravitational field will sort
the gas by specific entropy such that low-entropy gas lies inside
higher-entropy gas. We now address the question of how wide a spread
in specific entropy is required to prevent cooling producing a region
of outwards-decreasing specific entropy in which clouds could form. A
simple way of answering this question is to consider the cooling of
coronae that are initially polytropes of index $\gammap<\gamma$, and
we focus here on the Milky-Way--like models of \figref{figpoly}.

The right panels of \figref{KBfig} show the evolution of the Milky-Way
corona that starts from the polytrope with $\gammap=1.55$. The full
curve labelled ``$\sigma$'' in the bottom right panel shows this
corona initially has a significant entropy gradient in the outer
region, where an entropy depression formed in the adiabatic case. The
dotted curve for $\sigma$ in the bottom right panel shows that after
$1\Gyr$, the entropy gradient has been reversed, and there is again a
region of outwards-decreasing specific entropy. For polytropic indices
$\gammap\lta1.4$ a cooling catastrophe develops at the centre before
there has been significant cooling at the periphery, and for
$\gammap=1.4$ a point of inflection develops in $\sigma(M)$ at
$M=12.2\times10^{10}\msun$ around the time ($\approx0.5\Gyr$) when
cooling becomes catastrophically fast at the centre.

The central panels of \figref{KBfig} show that thermal instability
occurs only at radii in excess of $100\kpc$. This is the case because
the steep portion in the cooling curve, which drives the instability,
occurs at temperatures that are significantly below the central
temperature of the corona -- such temperatures are reached only close
to the edge of the corona because when the circular speed is assumed
constant (as here) they imply a steep density profile.  In fact, the
formation of a region of outwards-decreasing specific entropy depends
sensitively on the shape of the cooling function. When $\Lambda(T)$ is
that given by \cite{SutherlandD} for zero metallicity, no such region
forms because $\Lambda(T)$ does not fall steeply enough with
increasing temperature. At zero metallicity an initially adiabatic
corona develops an outwards-increasing entropy gradient on account of
the increase in the radiation rate with increasing density.

\begin{figure}
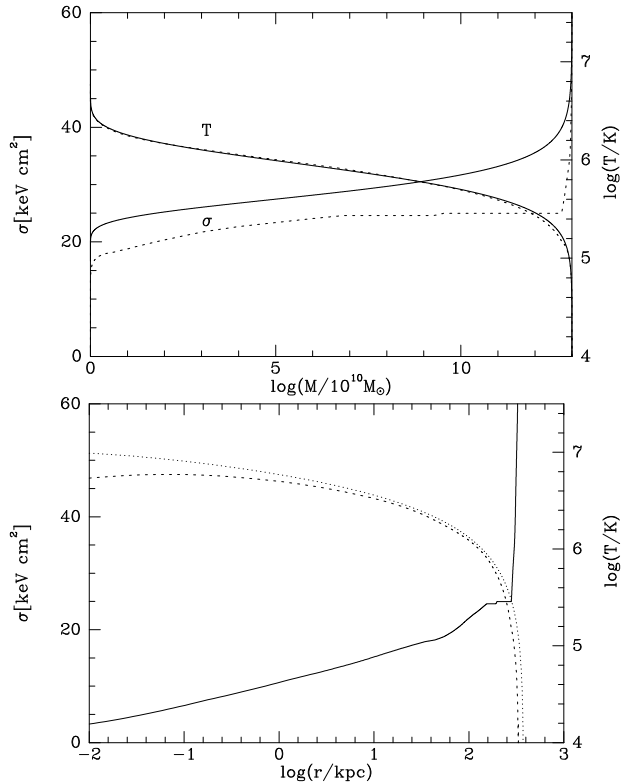

\centerline{\includegraphics[width=.96\hsize]{convectT.ps}}
\centerline{\includegraphics[width=.96\hsize]{convectr.ps}}
\caption{The evolution of the initial polytrope with $\gammap=1.55$ when
convection is assumed to prevent entropy decreasing
outwards. Top panel as a function of mass; lower panel as a function of
radius.}
\label{fig:convect}
\end{figure}

Once entropy is either independent of radius or outwards-decreasing,
there is nothing to inhibit bulk radial motion of gas.  In an
adiabatic atmosphere such motions need to be externally excited, for
example by tidal fields and the motion of dark-matter substructures
and satellite galaxies as they orbit through the host gravitational
potential. In a convectively unstable atmosphere, such motions will
arise spontaneously.  Radial bulk motions transfer heat between
different radial shells of a corona, and, as in the theory of stellar
structure, it is natural to assume that convective energy transport
will maintain an adiabatic entropy profile in any region in which
entropy would otherwise be outwards-decreasing.  Hence we modified the
code that produced \figref{KBfig} to include convection as
follows. After each timestep, the code looks for regions in which
entropy is outwards decreasing and sets the specific entropy
$s_i=\ln(\sigma)+\hbox{constant}$ of each mass shell in this region to
a common value $\overline{s}$. The search for regions of
outwards-decreasing entropy, and subsequent entropy adjustment is then
repeated until no such region is found. The entropy $\overline{s}$ is
reached as a result of the $i$th mass shell receiving an amount of
heat $\d q_i$ from the other shells, thus changing its specific
entropy by
 \begin{equation}\label{eq:dq}
\overline{s}-s_i=\d s_i={\d q_i\over T_iM_i},
\end{equation}
 where $T_i$ and $M_i$ are the temperature and mass of the $i$th shell. By
conservation of energy, $\sum_i\d q_i=0$, so after multiplying (\ref{eq:dq})
through by $T_iM_i$ and summing over $i$ we have
$\overline{s}=\sum_is_iT_iM_i/\sum_iT_iM_i$.

\figref{fig:convect} shows the evolution of the initially polytropic
corona plotted in the right panels of \figref{KBfig} when this
algorithm is used to simulate the effects of convection. One sees that
after a Gyr specific entropy is constant through nearly half the
mass. Thus metal-enhanced cooling combined with convection has
considerable potential for driving the outer half of a polytropic
corona to an adiabatic configuration, provided that the initial
entropy profile is shallow enough. However, the fact that polytropic
models are characterised by a finite boundary at which the temperature
drops to zero is crucial for these systems to evolve towards a
configuration with flat entropy profile in the outer regions. We do
not expect that these conditions are met in real galaxies, which are
surrounded by the intergalactic medium.

\section{Conclusions}

In this paper we examined the proposal that \HVC s of the Milky Way
and other disc galaxies form by condensation of the \WHIM\ through
thermal instability.  The linearised equations for the evolution of
thermal perturbations within a cooling, conducting, non-rotating
spherical corona show that unless specific entropy is a shallow
function of radius (a nearly adiabatic corona), regions in which the
temperature is initially anomalously low do not experience the
classical Field instability. In fact they are simply points at which
an internal gravity wave is momentarily in the ascendant, and after
half a Brunt-V\"ais\"al\"a period the material in this region will
sink to below its equilibrium radius and will then be an anomalously
warm region. Consequently, there is no thermal instability from which
clouds can form in a corona in which there is a significant gradient
in the specific entropy.

Clouds can form in a strictly adiabatic corona because in this case
buoyancy vanishes and the increase in the specific cooling function
$\Lambda(T)$ with decreasing temperature around $T\sim3\times10^5\K$
can cause the cooling rate to increase outwards notwithstanding the
steady decline in gas density with radius. This increase in the
cooling rate can cause the specific entropy to become
outwards-decreasing and thus the corona to become convectively
unstable. In these circumstances clouds of rapidly cooling material
will tumble inwards if thermal conductivity is too weak to carry
enough heat from the warm interior to the rapidly cooling periphery to
offset radiative cooling.

An examination of adiabatic coronae for three galaxies, NGC\,5746, the
Milky Way, and NGC\,6503 shows that thermal conductivity that is
suppressed to $1\%$ of Spitzer's fiducial value would eliminate
thermal instability entirely in NGC\,5746 and largely in the Milky
Way, where perturbations smaller than $10\kpc$ in extent could grow
only at $r\gta200\kpc$.  The efficiency with which thermal
conductivity can suppress thermal instability is greatest in the
hottest coronae because the Spitzer's conductivity is a strongly
increasing function of temperature. So in a low-mass galaxy such as
NGC\,6503, perturbations $10\kpc$ in extent could grow through most of
the corona, although a perturbation only a kpc in extent would be
stabilised almost everywhere.

It is implausible that all the gas in a corona initially has exactly
the same specific entropy. The gas will inevitably be spread over a
range of specific entropies, and within a few dynamical times
convection will sort the corona by specific entropy so that entropy is
outwards-increasing.  A convenient way of simulating the consequences
of such a spread in entropy is to consider coronae that are initially
polytropes with index $\gammap<\gamma$.  In such coronae there is a
significant radial gradient in specific entropy near the corona's
edge, where the temperature is low and thermal instability is most
likely. Non-linear simulations of the cooling of initially polytropic
coronae for the Milky Way show that when the polytropic index
$\gammap$ is smaller than $1.4$, cooling steepens the initial
outwards-increasing entropy gradient rather than reversing it and this
ensures that anomalously cool regions are buoyantly stabilised against
thermal instability.

Our conclusion that a flat entropy profile is a necessary condition
for the growth of thermal instability in galactic coronae is
consistent with the recent results of SPH simulations by
\cite{Kaufmann09}, who find that cold clouds form from hot gas only in
systems with flat entropy profiles. However, we find that except in
very low-mass systems, thermal conductivity \citep[which was not
considered by][]{Kaufmann09}, is capable of suppressing thermal
instability at most radii, even when the conductivity is strongly
suppressed.

In light of these results, is it likely that the clouds of \hi\ that
are actually observed have formed by cooling of coronal gas? A major
issue is that observed \hi\ clouds, such as Complex C and the filament
of NGC\,891 lie at radii $\sim10\kpc$ \citep{Oos07}.  Clouds are not
found as far out as the bounding radii of a nearly adiabatic
atmospheres that would contain the missing baryons, while theory
indicates that it is precisely near these radii that clouds may form
as a result of the coronal temperature falling through
$T\simeq3\times10^5\K$, where the cooling function is sufficiently
steep.  One might speculate that initially clouds have densities and
masses that lie below the thresholds for detection, and it is only
after they have fallen in to much smaller radii, where the confining
pressure is higher, and accreted coronal gas or merged with other
clouds, that they become detectable. In this case deeper surveys
should find clouds out to in excess of $100\kpc$. However, a
difficulty with this proposal is that small clouds are thermally
unstable only if the corona's thermal conductivity is heavily
suppressed -- in fact to less than a percent of Spitzer's value.

Thermal instability only occurs if the coronal temperature falls
through $T\simeq3\times10^5\K$. In the polytropic coronae discussed
here this condition is satisfied because $T$ vanishes at a finite
radius. However, this feature of a polytropic corona is artificial: in
reality the corona will at some point merge into the general
intergalactic medium, which is likely to have high entropy and a
temperature which could well be above $5\times10^5\K$.

Except in Section \ref{sec:timeevol} our analysis is linear in nature
and does not apply close to the star-forming disc since this region is
violently disturbed by expanding supershells. However, linear analysis
should be a reasonable approximation several kiloparsecs from the disc
where HVCs are observed.

A significant limitation of our work is that it excludes both magnetic
fields and rotation of the corona.  \cite{Loewenstein1990} and
\cite{Balbus91} showed that magnetic fields can in principle have a
big impact on the stability of cooling flows. However, the relevance
of the linear calculations presented by these authors to real cooling
flows and coronae is doubtful because in order to perform the analysis
one has to assume an unperturbed magnetic-field configuration that is
consistent with the unperturbed cooling flow -- this field turns out
to be radial. The modes obtained obviously reflect the particular
stress tensor of this field, and are valid only for perturbations that
do not significantly change the field geometry because $\delta{\bf
  B}\ll{\bf B}$. In cooling flows magnetic pressure is much smaller
than thermal pressure and probably on the order of the turbulent
pressure $\rho v_{\rm turb}^2$. Consequently the results of
\cite{Loewenstein1990} are valid only when the velocities associated
with modes are small compared to $v_{\rm turb}$. But we are precisely
interested in the turbulent motions themselves, so this condition will
be seriously violated.  Moreover, given that the magnetic pressure is
not larger than the turbulent pressure, the magnetic field will be
tangled by the turbulence rather than have the special geometry
required by the assumption that a steady state is maintained in the
presence of a radial accretion flow. It seems likely that in these
circumstances the safest procedure is simply to assume that the effect
of the (tangled) magnetic field is confined to augmenting the
isotropic pressure by $\sim B^2/2\mu_0$. In this case our results are
applicable.

Rotation, by contrast, may have a significant impact on thermal
stability.  Indeed, if clouds that condense from the corona are to
feed the growth of a centrifugally supported gas disc, rotation must
be dynamically important near the disc. However, clouds are observed
far from the disc, where rotation will be less important, and the
question is whether a level of rotation that is dynamically small (in
the sense that the centrifugal acceleration is much smaller than the
gravitational acceleration) can introduce thermal instability by
obstructing internal gravity waves. We reserve this difficult question
for a later paper.

We conclude that, if galactic haloes are well represented by
non-rotating, quasi-hydrostatic atmospheres, thermal instability is
not a viable mechanism for the formation of cold clouds in disc
galaxies as massive as, or more massive than, the Milky Way. It may
just work in low-mass systems if their (essentially unconstrained)
entropy profiles are almost perfectly flat. Our best guess is that the
observed population of \HVC s around the Milky Way and other large
nearby galaxies represent gas that has been stripped from satellites.
This conclusion leaves open the question of how gas moves from the
\WHIM\ to the disc, as it must if star formation is to be sustained.
\cite{FraternaliB08} present circumstantial evidence that the key to
this transfer is the galactic fountain, which constantly cycles disc
gas through the bottom of the corona. To understand this transfer we
need models of the corona that include significant centrifugal support
near the disc.


\label{lastpage}
\end{document}